\documentstyle[preprint,aps]{revtex}
\newtheorem{defn}{Definition}[section]
\begin{document}
\large
\title{Hamiltonian Dynamics of Darwin Systems}
\author{Alexander V. Shapovalov}
\address{Tomsk State University, Tomsk, RUSSIA\\
E-mail: shpv@phys.tsu.tomsk.su}
\author{Eugene V. Evdokimov}
\address{Research Institute of Biology and Biophysics, Tomsk, RUSSIA\\
E-mail: evd@biobase.tsu.tomsk.su}
\maketitle
\begin{abstract}
We present a Hamiltonian approach for the wellknown Eigen
model of the Darwin selection dynamics. Hamiltonization is carried out by
means of the embedding of the population variable space, describing
behavior of the system, into the space of doubled dimension by
introducing additional dynamic variables. Besides the study
of the formalism, we try to interpret its basic
elements (phase space, Hamiltonian, geometry of solutions) in terms
of the theoretical biology. A geometric treatment is given for the
considered system dynamics in terms of the geodesic flows in the
Euclidean space where the population variables serve as
curvilinear coordinates.
\par
The evolution of the
distribution function is found for arbitrary distributed initial
values of the population variables.
\end{abstract}

\section{Introduction} \label{Int}

Methods of Hamiltonian and Lagrangian analysis are extensively
applied during last decades in various fields of theoretical and
mathematical physics. These methods, however, have not been widely
used until now in mathematical theory of biological and ecological
systems which foundations originated from the works of the 10-30-es
by A.  Lotka,  V.  Volterra, J. Haldane and R. Fisher.
\par
The basic obstacle, by our view, is that the mathematical models
of biological subjects, being open systems in their essence, are
not supposed to be Hamiltonian ones from the outset.  On the other
hand, broad potentialities may appear for biological systems
to analyze some aspects (of both principle and technical character)  of their
dynamics, if the system admits Hamiltonization in a sense.
\par
In the present paper we suggest a Hamiltonian form for the wellknown
Eigen model of the Darwin selection dynamics.
To construct the Hamiltonian formalism, we extent the space of
population variables introducing additional degrees of freedom.
As a result the behavior of the system is described in terms of the
space with doubled dimension if compared to the original one.
Besides the study of the formalism, we try to interpret its basic
elements  (phase space, Hamiltonian, geometry of solutions) in terms
of the theoretical biology. A geometric treatment is given for the
considered system dynamics in terms of the geodesic flows in the
Euclidean space where the population variables serve as
curvilinear coordinates.
\par
The notion of a Darwin system (DS) as a formal object in
theoretical biology had been introduced by
M. Eigen to deduce laws of living system evolution from the
principles of theoretical physics and chemistry.
\par
By definition, the DS is an open system, which is constituted
of coupling units of different species (genotypes) self-copying
with a small number of errors
(convariantly self-reduplicating according to Timofeev-Resovskii
nomenclature \cite{Tim96}). The coupling units utilize a
substance and free energy of an external nutrient supply \cite{Eig79}.
\par
The  Darwin selection phenomenon  occurs in such systems
under the following necessary conditions:
the stability of system organization (i.e. the total
quantity of units of all species should conserve)
or the stability  of feed component influx.
Both of these constraints imply the flow through the system. A simple
experimental model of the DS (chemostat, turbidostat, etc.)
can be a system  where viruses and bacteria are defined as convariantly
self-reduplicating units (see, for example, \cite{Pe82}).
\par
The following system of differential equations proposed by M. Eigen
to describe the DS evolution:
\begin{equation} \label{1}
\dot x_i=x_i(A_iQ_i-D_i)+\sum_{j\ne i}^Nw_{ji}x_j-\Phi _i,
\end{equation}
here $i,j,k=1,\dots ,N$; $N$ is a species quantity of
self-reduplicating units (genotypes) in the system; $x_i$ is a specific
quantity (concentration) of the $i$-th species; $A_i$ ($D_i$) is a specific
reproduction (death) velocity of the $i$-th species; $Q_i$ is a parameter of
the reduplication quality of the $i$-th species (takes the values from $0$
to $1$); $w_{ji}$ is a specific velocity of the error reduplication of the
$j-$th species into the $i-$th one; $\Phi _i$ is a dilution parameter which
usually equals to $D x_i$. Here $D$ is a dilution flow rate.
\par
Further, as it has been shown by M. Eigen \cite{Eig79}, the system
(\ref{1}) can be put in a  simpler form
\begin{equation} \label{2}
\dot y_i=y_i(\mu _i({\bf S})-D)
\end{equation}
using obvious relations between $A_i$, $Q_i$, $w_{ji}$ and by introducing
new population variables $(y_i)$, being related to the so called
"quasispecies" which are described by certain combinations of original
variables $(x_i)$.
Here $i,j,k=1,\dots ,N$; $N$ is a (constant) quasispecies quantity in
the system; $y_i$ has a meaning of a specific quantity (concentration) of
the $i$-th quasispecies; ${\bf S}=(S_1,\dots ,S_f)$ are the concentrations
of the external nutrient supply components; $\mu _i({\bf S})$ is a
generalized Maltuzian parameter which has a meaning of the reproduction
specific velocity of the $i$-th quasispecies and depends on ${\bf S}$.
The $\mu _i({\bf S})$ is an algebraic combination of all $A_k,D_k,w_{jk}$ in
which $A_i$ makes a major contribution. The $\mu _i({\bf S}),D$ have
dimension of inverse time. The "quasispecies" introduced by M. Eigen
as new population variables have a clear biological sense of clones,
systems of organisms with a prevalence of the determined genotype
in the system and  small admixtures  of some other genotypes
connected with the leading
one by the mutation and recombination transitions \cite{Eig79}.
\par
There are two commonly recognized types of DS under the constraints
imposed above on the system (\ref{1}). Recall the definitions of
these types.
\begin{defn} \label{def1}
{\sl A Darwin system is referred to as DS with stable organization (DSSO)
if} $\sum_{i=1}^Ny_i$ {and ${\bf S}$ are constant. A Darwin system with}
$D={\rm const}$ {is referred to as DS with a stable flow (DSSF)}.
\end{defn}
It is known that the representation of the system (\ref{2}) in terms
of relative variables having a meaning of shares (probabilities) of the
quantities \cite{Ebel82} transforms (\ref{2}) into the classical form of the
Fisher system of equations which describes the Darwin selection dynamics in
the panmictic populations under the special requirement for the
fitness function \cite{Fish30}.
The system (\ref{2}) is also valid for description
of interspecies competition for the grow-determining substrate
when there is no migration. In view of significance of the systems
(\ref{1}),(\ref{2}) in biology and ecology they are investigated in
a large number of works since the paper \cite{Fish30} has appeared.
So we touch upon some of these publications relevant for the
present paper.
\par
An exact solution of (\ref{2}) is known only for
$\mu _i={\rm const}$ \cite{Zog70}, a general form of
approximate solution is found
by Jones \cite{Jon79} for the $\mu _i$ changing in time. A general review
of the properties of system (\ref{2}) in the Fisher modification is given in
\cite{Log78}, \cite{Svir82}. Stability of the stationary solutions of (\ref{2})
is analyzed in detail by Pykh \cite{Pyh83}. Feistel and Ebeling have studied
behavior of the Eigen-Fisher system  on the adaptive Right's landscape
and shown that in a number of cases the system can be represented
in the canonical gradient form \cite{Ebel82}, \cite{Ebel84}.
Isida \cite{Isida84} investigated non-equilibrium thermodynamics of the systems
described by the Eigen equations.
As is known, there are two different approaches to describe Darwin, Fisher
and reducible to them Volterra systems by means of extremal principles.
The first (teleological) approach assumes to extremalize of a functional
having a certain biological sense as a finite aim of the evolution.
For example, it might be maximum of an average population fitness
\cite{Fish30}, or maximum of an average productivity \cite{Eig79}, or
maximum of a  reproductive potential \cite{Log78} etc. The second
approach assumes to achieve  the extremum of some function
of dynamic variables in each moment of time. Thus, in reference
\cite{Evd95} it is shown that the requirement for Kulback's measure
to have a maximum value leads to the relations
which are the solutions of (\ref{2}) under the special conditions
imposed on the system in the approximation of the DSSO.
As a next example we refer to the work
\cite{Svir82} where variational form of the Eigen-Fisher
dynamics in a special case
is built up by introduction of a metrics of a Riemann space. Despite
the fact that the dynamics of the Eigen-Fisher type systems are managed to
represent in the form of extremal principles in some special cases, the
problem stays far from complete solution as it has been explained in
\cite{Log78}.
\par
The Hamiltonian form of the DS dynamic equations, being proposed in the
present work, leads to the conventional variational principle,
which is habitual in mechanics.

\section{DS with a stable organization}  \label{II}

In accordance with def.\ref{def1}, Eqs. (\ref{2}) assume the following
form  for the DSSO case:
\begin{equation}
\label{3}\dot y_i=y_i\left(\mu_i-{\sum _{j=1}^Ny_j\mu_j}
\biggm/{\sum _{j=1}^Ny_j}\right), \quad \mu_i={\rm const}.
\end{equation}
The dimensionless variables $p_i=y_i/\sum _{j=1}^Ny_j$ have a meaning of
shares (probabilities) of the quantities.
Taking into account that $\sum _{j=1}^N y_j = {\rm const}$, we
derive from (\ref{2}) for $p_i$  the equations:
\begin{equation}
\label{4}\dot p_i=p_i(\mu_i-{\sum _{j=1}^Np_j\mu_j}).
\end{equation}

If $p_N$ is eliminated solving the constraint $\sum _{j=1}^Np_j=1$,
one can rewrite (\ref{4}) as follows:
\begin{equation}
\label{5}\dot p_i=p_i(a_i-{\sum _{j=1}^{N-1}p_ja_j}).
\end{equation}
Here and up to the end of this section we assume:
$i,j,k=1,\dots ,N-1$,\,\,$a_i=\mu_i - \mu_N$, $\mu_N$
is a Maltuzian parameter of $N$-th quasispecies
taken as a gauge one, for example, by the criterium of minimal value of
$\mu_i$. Next, to simplify (\ref{5}) we introduce the following variables:
\begin{equation}
\label{6}
z_i= \log [p_i/(1-\sum _{j=1}^{N-1}p_j)].
\end{equation}
Accounting the condition $1-\sum _{j=1}^{N-1}p_j = p_N$, the new
variables $z_i$  might be called as informational ones,
inasmuch as they reflect, in a spirit of the  Shannon's theory
\cite{Shan49}, a quantity of information for every independent degree of
freedom co-ordinated to the eliminated $N$-th variable.
\par
Consider variables (\ref{6}) as coordinates of a manifold ${\cal A}$
which naturally can be named as an information one. A state of the system
in ${\cal A}$ is characterized by a point
${\bf z}=$ $(z_1,z_2,\dots ,z_{N-1})$. The system dynamics
${\bf z}(t)$ is described by a velocity vector $\dot {{\bf z}}(t)$.
The system (\ref{5}) in the variables ${\bf z}$ takes
the simple form,
\begin{equation}\label{7}
\dot z_i=a_i,
\end{equation}
and has the general solution
\begin{equation} \label{7aa}
z_i(t)=z_{i_0}+a_i t.
\end{equation}
Hence, evolution of the DSSO in the "information" variables (\ref{6}) is a
uniform rectilinear motion. Taking "mass" to be a unit, we can conceive that
the $a_i$ is the $i$-th component of the momentum.
In accordance with Hamiltonian formalism, we will interpret $(z_i)$
and $(a_i)$ as the position and momentum variables, respectively,
in a $2(N-1)-$dimensional phase space ${\cal M}$ of the system.
Introducing  Hamiltonian of the system
\begin{eqnarray*}
h=\frac 12 \sum _{j=1}^{N-1} a_j^2,
\end{eqnarray*}
we can write down the Hamiltonian form of the system (\ref{7}) as follows:
\begin{equation}\label{9}
\left\{
\begin{array}{rl}
& \dot z_i=\partial h/\partial a_i=a_i, \\
& \dot a_i=\,-\partial h/\partial z_i=0.
\end{array}
\right.
\end{equation}
In the framework of the considered Hamiltonian approach we can state the
following variational principle for the DSSO dynamics.
Lagrange function of the system (\ref{9}),
$l=$ $\sum_{j=1}^{N-1}a_j\dot z_j-$ $h$, after  the momentum variables
$(a_i)$ eliminating, takes the form:
\begin{equation} \label{10}
l(z,\dot z)=\frac 12\sum_{j=1}^{N-1}\dot z_j^2.
\end{equation}
The minimal action principle is:
\begin{equation} \label{11}
A=\int_0^tl(z,\dot z)(\tau )d\tau ,\qquad \delta A=0.
\end{equation}
The variational principle (\ref{10}),(\ref{11}) leads to the following
treatment of the DS dynamics:
\par
\noindent a DSSO evolutes in the information manifold  ${\cal A}$
in such a way which minimalizes the squared norm of the respective
velocity of variation of information in the system at the considered
time interval.
\par
 Canonical form of the dynamic equations (\ref{5}) is found from the
canonical transformation $(z_i,a_i)$ $\rightarrow $ $(p_i,r_i)$, where $r_i$
are additional variables canonically conjugate to $p_i$. This transformation
is:
\begin{equation} \label{12}
\left\{
\begin{array}{rl}
& p_i=e^{z_i}/(1+\Delta ),\quad \Delta \equiv \sum_{j=1}^{N-1}e^{z_j}, \\
\,\, & r_i=(1+\Delta )[a_ie^{-z_i}+\sum_{j=1}^{N-1}a_j].
\end{array}
\right.
\end{equation}
The inverse transformation:
\begin{equation} \label{13}
\left\{
\begin{array}{rl}
& e^{z_i}=p_i/(1-\Omega ),\qquad \Omega \equiv \sum_{j=1}^{N-1}p_j, \\
& a_i=p_i(r_i-w),\quad w=\sum_{j=1}^{N-1}r_jp_j.
\end{array}
\right.
\end{equation}
Eqs. (\ref{9}),(\ref{12}),(\ref{13}) lead to the original system (\ref{5}) to
be supplemented by the equations for $r_i$:
\begin{equation} \label{15}
\left\{
\begin{array}{rl}
& \dot p_i=p_i[p_i(r_i-w)-\sum_{j=1}^{N-1}p_j^2(r_j-w)], \\
& \dot r_i=r_i\sum_{j=1}^{N-1}p_j^2(r_j-w)-p_i(r_i-w)^2.
\end{array}
\right.
\end{equation}
Using (\ref{7aa}), (\ref{12}),(\ref{13}) we obtain the solution of
(\ref{15}) in the form:
\begin{equation} \label{15a}
p_i(t)=p_{0i}\frac{\exp [p_{0i}(r_{0i}-w_0)t]}{1-\Omega _0+
\displaystyle\sum_{k=1}^{N-1}p_{0k}\exp [p_{0k}(r_{0k}-w_0)t]},
\end{equation}
\begin{equation}\label{15b}
\begin{array}{ll}
& r_i(t)=\left\{ r_{0i}\exp (-p_{0i}(r_{0i}-w_0)t)+w_0[1-
\exp (-p_{0i}(r_{0i}-w_0)t)]\right\} \times \\
& \left\{ 1-\Omega _0+\sum_{k=1}^{N-1}p_{0k}
\exp (p_{0k}(r_{0k}-w_0)t)\right\} .
\end{array}
\end{equation}
Here $p_{0k}=p_k(t)|_{t=0}$, $r_{0k}=r_k(t)|_{t=0}$ are the initial
conditions for the system (\ref{15}), $w_0=\sum_{k=1}^{N-1}p_{0k}r_{0k}$,
$\Omega _0=\sum_{k=1}^{N-1}p_{0k}$. Expressions $r_0(t,p,r)$, $p_0(t,p,r)$
are obtained from (\ref{15a}), (\ref{15b}) by the substitution:
$p_{0i}\leftrightarrow p_i$, $r_{0i}\leftrightarrow r_i$, $t\rightarrow -t$.
The system (\ref{15}) is the Hamiltonian one in terms of the canonical
coordinates $(p_i,r_i)$ ($p_i$ are the position variables and $r_i$ are
the momentum ones) with the Hamiltonian
\begin{equation} \label{15c}
h=\frac 12\sum_{j=1}^{N-1}p_j^2(r_j-w)^2.
\end{equation}
The respective dynamics can be represented in terms of the geodesic
flows of the $(N-1)-$
dimensional Euclidean space where $p_i$ serves as curvilinear coordinates
and $z_i$ are the Cartesian ones. To impart tensor nature to the notations,
let $z_i\rightarrow z^i$, $p_i\rightarrow p^i$ leaving subscripts at $a_i$,
$r_i$. The Hamilton function (\ref{15c}) can be rewritten as
\begin{equation} \label{15d}
h=\frac 12\sum_{j=1}^{N-1}g^{ij}(p)r_ir_j.
\end{equation}
Here $g^{ij}(p)$ are the contravariant components of the metric tensor in
the coordinates $(p_i)$:
\begin{equation} \label{15e}
\begin{array}{ll}
g^{ij}(p)=\displaystyle \sum_{k=1}^{N-1}\displaystyle\frac{\partial {p^i}}
{\partial {z^k}}\frac{\partial {p^j}}{\partial {z^k}}=(p^i)^2\delta
^{ij}-[(p^i)^2p^j+p^i(p^j)^2]+R^2p^ip^j,
\end{array}
\end{equation}
where $R^2=\displaystyle\sum_{k=1}^{N-1}(p^i)^2$. The covariant components
are:
\begin{equation} \label{15g}
\begin{array}{ll}
g_{ij}(p)=\displaystyle \sum_{k=1}^{N-1}\displaystyle\frac{\partial {z^k}}
{\partial {p^i}}\frac{\partial {z^k}}{\partial {p^j}}=
\displaystyle \frac{\delta _{ij}}{(p^i)^2}+
\displaystyle \frac 1{p^i(1-\Omega )}+\displaystyle
\frac 1{p^j(1-\Omega )}+\displaystyle \frac{N-1}{(1-\Omega )^2},
&  \\
&  \\
\det (g^{ij}(p))=(1-\Omega )^2\Pi _{j=1}^{N-1}(p^j)^2,
\end{array}
\end{equation}
$\sum_{k=1}^{N-1}g^{ik}g_{kj}=\delta _j^i$, \thinspace \thinspace
\thinspace ($\delta _{ij},\delta ^{ij},\delta _j^i$ are the Kronecker
deltas). The Lagrange function,
\begin{displaymath}
l=\sum_{k=1}^{N-1}\dot p^kr_k-h=
\frac 12\sum_{k,l=1}^{N-1}g_{kl}(p)\dot p^k\dot p^l,
\end{displaymath}
defines the Euler-Lagrange equations,
\begin{equation} \label{E-L}
{d}/{dt}({\partial l}/{\partial \dot p^j})-{\partial l}/{\partial p^j}=0,
\end{equation}
which take the form of geodesic equations:
\begin{displaymath}
\displaystyle \frac{d^2p^j}{dt^2}+
\sum_{k,l=1}^{N-1}\Gamma _{kl}^j(p)\dot p^k \dot p^l=0,
\end{displaymath}
where $\Gamma _{kl}^j(p)=\displaystyle
\frac 12\sum_{s=1}^{N-1}g^{js}({\partial g_{ks}}/{\partial {p^l}}+{\partial
g_{ls}}/{\partial {p^k}}-{\partial g_{kl}}/{\partial {p^s}})$ are the
Cristoffel symbols.
\par
Let us point out that the coordinates $p_i$ are nonorthogonal ones
due to nondiagonality of the metrics in these coordinates.
\par
The DS trajectories $p_i(t)$ are usually mapped  as curved lines
in the coordinates $p_i$ supposing them to be orthogonal Cartesian ones
(see, for example,\cite{Eig79}), whereas this motion can be represented as
straight lines in the Cartesian coordinates $z_i$. Twisting of the
trajectory with respect to $(p_i)$ is the result of the curvilinearity
of these  coordinates. The existence of the coordinates ($z_i$), where
the DSSO dynamics is represented by a uniform rectilinear motion,
illustrates  the absence of a coupling between
quasispecies and  any external factors in the course of the
Darwin selection process. From this
standpoint, the DSSO can serve as a biology-theoretic
analogue of an inertial reference system in the Newtonian mechanics.
\par
Let us note  that the $a_i$ give rise to an Abelian algebra of
integrals of the system (\ref{15}) which is integrable by Liouville with
respect to the conventional Poisson brackets $\{f,g\}=$
$\sum _{j=1}^{N-1}(\partial f/\partial p_j
\partial g/\partial r_j-$ $\partial
g/\partial p_j\partial f/\partial r_j)$. The system
\begin{equation} \label{16}
\left\{
\begin{array}{rl}
& dp_i/d\alpha=\partial a_q/\partial r_i=p_q(\delta _{qi}-p_i), \\
& dr_i/d\alpha=-\partial a_q/\partial p_i=-\delta _{iq}(r_q-w)+p_qr_i
\end{array}
\right.
\end{equation}
defines a flow  generated  by  $a_q$  in the phase space ${\cal M}$
which leaves
Eqs. (\ref{15}) invariant, $\alpha $ is a group parameter. Integration
gives:
\begin{equation} \label{17}
\begin{array}{rl}
& p_q(\alpha)=p_{q}(0)e^\alpha V(\alpha)^{-1}; \\
& p_i(\alpha)= p_{i}(0)V(\alpha)^{-1}, \quad i\ne q ; \\
& r_q(\alpha)=e^{-\alpha}V(\alpha)^{2}[r_q(0)+W(e^{\alpha}-1)V(\alpha)^{-1}]; \\
& r_i(\alpha)= r_i(0)V(\alpha), \quad i\ne q.
\end{array}
\end{equation}
\par
Here $V(\alpha)\equiv 1+(e^\alpha -1)p_q(0)$,
$W\equiv \sum _{j\ne q}r_j(0)p_{j}(0)$, $p_{i}(0), r_i(0)$
are initial conditions for
(\ref{16}). Let us note, that  the variables $p_i$ in (\ref{17}) transform
independently of $r_i$ so that one can directly use (\ref{17}) for the
original system (\ref{5}).
\par
In conclusion of this section we consider DSSO dynamics with less number of
constraints then in the def.\ref{def1}. Let us retire only the
requirement of constant of total quantity of all species in the
system and remove the restrictions on the nutrient supply concentration
and on some other parameters of the medium, for example, on a temperature.
This leads for  $\mu _i$ to be certain functions of time.
Since  $\mu _i$ are not involved in the transformation from the
original variables to the  "information" ones, Hamiltonian form of
the DS dynamics can be thereby constructed  in the considered case too.
\par
Let $f_i $ be the first derivative of $a_i$, $\dot a_i=f_i$. Since
$a_i$ serve as momentum variables in the DS dynamics, then the derivatives of
$f_i $ can be considered as components of a force. It is  easy to show that
the Hamiltonian of the considered system takes the form:
\begin{equation} \label{18}
h(z,a)=\sum _{i=1}^N(\frac 12 a_i^2-z_if_i).
\end{equation}
Respectively, Hamiltonian form of Eqs. (\ref{9}) is:
\begin{equation} \label{19}
\left\{
\begin{array}{rl}
& \dot z_i=\partial h/\partial a_i=a_i, \\  & \dot a_i=
-\partial h/\partial z_i=f_i.
\end{array}
\right.
\end{equation}
The Hamiltonian (\ref{18}) plays a role of an "energy" in the Darwin
dynamics if functions $f_i$ are constant in time. The second term in
(\ref{18}) has a sense of a "potential energy". The DS
dynamics  described by the system (\ref{19}) is  a motion in the field
of the "force" $f_i$. The additional "potential" term leads to the
variational principle (\ref{10}),(\ref{11}) with the
modified Lagrangian:
\begin{equation}\label{20}
l=\sum _{j=1}^{N}\dot z_ia_i -h=\sum_{j=1}^{N}(\frac 12 \dot z_i^2+z_jf_j).
\end{equation}

\section{DS with a stable flow} \label{III}

To provide Eqs. (\ref{1}) to be a complete system in the case of DSSF
(in the sense of def.\ref{def1}), one has to define explicitly the
functions $\mu_i({\bf S})$ and to extend the system introducing
equations for ${\bf S}$. Here and further $i,j,k=1,\dots ,N$.
\par
The wellknown Monod function \cite{Pirt78} is commonly recognized
for DSs. It depends on one independent variable $S$ which
has a meaning of concentration of the substrate being in
physiological minimum (growth limiting  substrate). That is
the concentration vector ${\bf S}$ is reduced to one component
$S$.
The dynamic system in this case takes the form:
\begin{equation}\label{1a}
\begin{array}{ll}
\dot y_i=y_i (\mu_i(S)-D), \\
\dot S=D(S_0-S)-{\sum _{j=1}^Ny_j\mu_j},\\
\mu_i= m_i S/(K_i+S),
\end{array}
\end{equation}
where $m_i$ is a maximal reproduction specific velocity,
$K_i$ is a saturation parameter of  $i$-th quasispecies,
$m_i, K_i={\rm const}$, $S$ is a concentration of the growth limiting
substrate in the system, $S_0$ is its concentration
in the external flow.
\par
There is no apparent way to solve the system (\ref{1a}) and to hamiltonize
it under arbitrary values of parameters.
\par
For the further analysis of this system let us take into account some
qualitative peculiarities of its dynamics which are known from experimental
data and  numerical simulations \cite{Dyk81}, \cite{Dyk93}.
\par
The system dynamics in general case is characterized by
two basic stadium:
\par
\noindent the first one is an initial transient with a period
of order $2/D\div 4/D$ where the function $S(t)$ varies
strongly and nonlinearly;
\par
\noindent the second one is a quasistationary behavior
in the interval $4/D<t<\tau$  where $\tau$ is an observation
time of the system.  $S(t)$ can be approximated in this case by a linear
function in this range up to 3-7\%.  Let
\begin{eqnarray*}
S(t)=\bar S+b(t-\frac 12\tau ).
\end{eqnarray*}
Here $\bar S$ is an average value of the $S(t)$ during the observation
time $\tau$ of the system, $b$ is an average velocity of the
concentration variation of the rate-determining substrate
in the system.
\par
Moreover, $\mu_i(S)$  can be also approximated by a linear
function in the range of the quasistationary behavior
where  $S(t)<K_i$.
\par
Expanding Monod function (\ref{1a}) in a neighborhood of the
point $S=\bar S$, we have in the second order:
\begin{eqnarray*}
\begin{array}{rl}
       &\mu_i(S)=\tilde \alpha _i+\tilde \beta_i S,\\
       &\tilde \alpha_i=\mu_i(\bar S)-
\bar S(\partial \mu_i(\bar S)/\partial \bar S)=
m_i\bar S^2/(K_i+\bar S)^2, \\
        &\tilde \beta _i=\partial \mu_i(\bar S)/\partial \bar S=
m_i K_i/(K_i+\bar S)^2.
\end{array}
\end{eqnarray*}
\par
With the approximations above, the DSSF dynamics in the quasistationary
stadium is described by the following equations:
\begin{equation}\label{4a}
\dot y_i=y_i(\alpha _i+\beta _i\cdot t),
\end{equation}
where $\alpha _i=\tilde \alpha _i$
$+\tilde \beta _i(\bar S-\frac 12 b\tau)-D$,
$\beta_i=$$b\cdot \tilde \beta _i$.
It is easily  to give a Hamiltonian form for the system (\ref{4a}).
Choosing canonical variables in a $2N-$dimensional phase space as
\begin{eqnarray*}
z_i=\log y_i, \quad a_i=\alpha _i+\beta _i\cdot t
\end{eqnarray*}
and the Hamiltonian
\begin{eqnarray*}
h(z,a)=\sum _{i=1}^N(\frac 12 a_i^2-z_i\beta _i),
\end{eqnarray*}
we arrive to the Hamiltonian form of Eqs. (\ref{4a}):
\begin{equation}\label{7a}
\left\{
\begin{array}{rl}
       &\dot z_i=\partial h/\partial a_i=a_i,\\
       &\dot a_i= -\partial h/\partial z_i=\beta _i.
\end{array} \right.
\end{equation}
Note, that the hamiltonian form  of the DSSF (\ref{7a}) is similar to the
DSSO case (\ref{19}) and  it describes uniformly accelerated rectilinear
motion in  space of variables  $(z_i)$ in a constant
and homogeneous external field with the  potential
$U(z)=-\sum _{j=1}^{N}\beta _j z_j$ determining a constant
"force" ${\bf F}={\bf \beta}$.
The Lagrange function is:
\begin{equation} \label{8a}
l=\sum _{j=1}^{N}\dot z_ia_i -h=
\sum_{j=1}^{N}(\frac 12 \dot z_i^2+z_j\beta _j).
\end{equation}
The variational principle has the form (\ref{11}) where  $l(z,\dot z)$
is of the form (\ref{8a}).
Canonical transformation  $(z_i,a_i)\rightarrow (y_i,r_i)$,
\begin{equation}\label{9a}
\left\{
\begin{array}{rl}
       &z_i=\log y_i,    \\
       &a_i= y_i r_i,
\end{array} \right.
\end{equation}
allows to rewrite  (\ref{4a}) in Hamiltonian form,
\begin{equation}\label{10a}
\left\{
\begin{array}{rl}
       &\dot y_i=\partial h/\partial r_i=y_i^2 r_i,\\
       &\dot r_i= -\partial h/\partial y_i=-y_ir_i^2+\beta _i y_i^{-1},
\end{array} \right.
\end{equation}
with the Hamiltonian
\begin{eqnarray*}
h(y,r)=\sum _{i=1}^N(\frac 12 y_i^2 r_i^2 -\beta _i\log y_i)=
\sum _{i,j=1}^N\frac 12 g^{ij}(y)r_i r_j + U(y),
\end{eqnarray*}
$U(y)=-\sum _{i=1}^N \beta _i\log y_i$.
The solution of the system  (\ref{10a})
with the initial conditions  $y_i(t)|_{t=0}=y_{i0}$,
$r_i(t)|_{t=0}=r_{i0}$ takes the form :
\begin{equation}\label{4aa}
\begin{array}{ll}
      &y_i(t)=y_{i0}\exp (y_{i0}r_{i0}t+\frac 12\beta _it^2),    \\
      &r_i(t)=\displaystyle \frac { y_{i0} r_{i0}+
\beta _i t}{y_{i0}\exp(y_{i0}r_{i0}t+\frac 12 \beta _it^2)}.
\end{array}
\end{equation}
\par
Expressions of  $y_0(t,y,r)$, $r_0(t,y,r)$ as in the case of
(\ref{15a}), (\ref{15b}) are obtained from (\ref{4aa})
by substitution: $y_{0i}\leftrightarrow y_{i}$,
$r_{0i}\leftrightarrow r_{i}$, $t\rightarrow -t$.
\par
Redenoting $y_i\rightarrow y^i$, we can write down
contravariant and covariant components of the metrics as follows:
$g^{ij}(y)=\delta ^{ij}(y^i)^2$,
$g_{ij}(y)=\delta _{ij}(y^i)^{-2}$, respectively.
The Lagrange function,  $l=\sum _{k=1}^{N} \dot y^k r_k -h$,
and the Euler-Lagrange equations (\ref{E-L})
take here the form:
\begin{displaymath}
l= \frac 12\sum _{i,j=1}^{N} g_{ij}(y) \dot y^i\dot y^j -U(y),
\end{displaymath}
\begin{displaymath}
\displaystyle \frac {d^2 y^j}{dt^2}+
\sum _{k,l=1}^{N} \Gamma ^j_{kl}(y)\dot y^k\dot y^l=
-\sum _{k=1}^{N} g^{jk}(y)\displaystyle \frac {\partial U(y)}{\partial y^k},
\end{displaymath}
respectively,
where $\Gamma ^j_{kl}(y)$ are the Christoffel symbols related to the
metrics  $g_{ij}(y)$.

\section{Distribution function of initial data for DS}\label{IV}

Hamiltonian form of the DS dynamics allows one to investigate
evolution of an initial data distribution.
Consider this problem for the DSSO case in more detail.
Let initial data for the DSSO are given in a domain
${\cal O}$ of phase space ${\cal M}$ with a distribution function
$f_0(p,r)$, $f_0(p,r)\ge 0$, $\int _{\cal O} f_0(p,r)dp dr=1$.
The problem of initial data indeterminancy arises for the Darwin
systems due to natural restrictions on the exactness of measurements
and Poisson character of initial values distribution of separate
genotypes quantity.
Consider the evolution of the distribution function $f(t,p,r)$
under the condition that
\begin{equation} \label{1b}
f(t,p,r)|_{t=0}=f_0(p,r).
\end{equation}
Supposing that the randomness is brought in the DS behavior only
in the form of random distribution of initial data and further
DS evolution is determined and is subjected to the Hamiltonian
equations above, we deduce that the distribution remains constant
on the phase trajectories on account of conservative character
of the Hamiltonian
dynamics. In other words, the distribution function $f(t,p,r)$
obeys the Liouville equation:
\begin{equation} \label{2b}
\displaystyle \frac {\partial f}{\partial t}+\{f,h\}=0.
\end{equation}
The solution of (\ref{2b}) with the initial condition (\ref{1b})
is obtained by substitution of the expressions $p_0(t,p,r)$, $r_0(t,p,r)$
inverting formulas $(\ref{15a})$, $(\ref{15b})$ into the
initial function $f_0$:
\begin{equation} \label{3b}
f(t,p,r)=f_0(p_0(t,p,r), r_0(t,p,r)).
\end{equation}
As an illustration let us consider more specifically the
simplest case of phase space $N=2$ when a point of phase space ${\cal M}$
is defined by one  coordinate ($p$) and one momentum ($r$) variables.
In this case   $p_0(t,p,r)$, $r_0(t,p,r)$ are
written down in the form:
\begin{eqnarray*}
p_0(t,p,r)=\frac 12[1-\tanh (t\cdot \theta +\delta)], \\
r_0(t,p,r)=8\theta \cosh ^2(t\cdot \theta+\delta),
\end{eqnarray*}
where $\theta =\frac 12 rp(1-p)$, $\delta =
\frac 12 \log \displaystyle \frac {(1-p)}{p}$.
\begin{equation}\label{4b}
f(t,p,r)=f_0(\frac 12[1-\tanh (t\cdot \theta +\delta)],
8\theta \cosh ^2(t\cdot \theta+\delta) ).
\end{equation}
Eq. (\ref{4b}) allows one to obtain time-dependent distribution
function of $p$:
\begin{eqnarray*}
f(t,p)=\int f(t,p,r) dr.
\end{eqnarray*}
Evolution of the distribution function can be inferred from
the transformation of equiprobabilistic surfaces  which
are obtained from equation $f(t,p,r)={\rm const}$
in different moments of time  $t$.
\par
Similar results can be easily obtained  in the case of DSSF.
Evolution of the distribution function is given by (\ref{3b})
in which it is necessary to substitute the equations
$y_0(t,y,r)$, $r_0(t,y,r)$ inverting (\ref{4aa}).

\section{Conclusion}\label{VI}

Hamiltonian form of the dynamic equations describing
Darwin selection process, being demonstrated here by the simple
examples of the DSs, leads to an interesting,  by our opinion,
aspects of the dynamics. For example, DS can be investigated
in the framework of near-integrable Hamiltonian systems with
more complicated functions $\mu_i$, $D$  then studied in the
present work.
\par
It is pertinent to note that in the present work the phase
space ${\cal M}$ includes original information space ${\cal A}$.
Such way of Hamiltonization seems to be more preferable
if compare to one proposed in \cite{Cron94} for the
Volterra-Lotka type systems where  phase space is built up
from the original variables. In the latter case a number of constraints
are imposed on the system that restricts the generality.

\end{document}